 \documentclass[final,3p,times,usenatbib]{elsarticle}

\usepackage[english]{babel}
\usepackage[latin1]{inputenc}
\usepackage{amsfonts} 
\usepackage{amsmath}
\usepackage{amssymb}
\usepackage{amsthm}
\usepackage{enumerate}
\usepackage{hyperref} 
\usepackage{graphicx} 
\usepackage[usenames,dvipsnames]{color}
\usepackage{rotating}
\usepackage{epsfig} 
\usepackage{graphicx}
\usepackage{threeparttable}
\usepackage{multirow}
\usepackage{url}
\usepackage{booktabs} 
\usepackage{lineno}

\usepackage{natbib}
\setcitestyle{authoryear,open={((},close={))}}

\newtheorem{definition}{Definition}[section]

\newtheorem{theorem}[definition]{Theorem}
\newtheorem{lemma}[definition]{Lemma}

\journal{Journal  }

\begin{document}

\begin{frontmatter}


\title{A class of two-sample nonparametric statistics for binary and  time-to-event outcomes}



\author{Marta Bofill Roig}
\ead{marta.bofill.roig@upc.edu}
\author{Guadalupe G\'omez Melis}  
\ead{lupe.gomez@upc.edu}

\address{Departament d'Estad\'{i}stica i Investigaci\'{o} Operativa, Universitat Polit\`{e}cnica  de Catalunya, Barcelona, Spain}

\begin{abstract} 
We propose a class of two-sample statistics for testing the equality of proportions and the equality of survival functions. We build our proposal on a weighted combination of a score test for the difference in proportions and  a Weighted Kaplan-Meier statistic-based test for the difference of survival functions. The proposed statistics are fully non-parametric and do not rely on the proportional hazards assumption for the survival outcome. We present the asymptotic distribution of these statistics, propose a variance estimator and show their asymptotic properties under fixed and local alternatives. We discuss different choices of weights including those that control the relative relevance of each outcome and emphasize the type of difference to be detected in the survival outcome. We evaluate the  performance of these statistics with a simulation study, and illustrate their use with a randomized phase III cancer vaccine trial. We have implemented the proposed statistics in the R package \texttt{SurvBin}, available on GitHub (https://github.com/MartaBofillRoig/SurvBin).
\end{abstract}


\end{frontmatter}



\section{Introduction} 

In many clinical studies, two or more endpoints are investigated as co-primary aiming to provide a comprehensive picture of the treatment's benefits and harms.  
Survival analysis  has often been the sharp focus of clinical trial research. However, when there is more than one event of interest, the time until the appearance of the event is not always the unique center of attention; often the occurrence of an event over a fixed time period is as well an outcome of interest.

In the context of cancer immunotherapies trials, short-term binary endpoints based on the tumor size, such as objective response, are common in early-phase trials, whereas overall survival remains the gold standard in late-phase trials (\citealp{Wilson2015,Ananthakrishnan2013}). 
Since traditional oncology endpoints  may not capture the clinical benefit of cancer immunotherapies,  the idea of looking at both tumor response and  survival has grown from the belief that together may achieve a better characterization of the clinical response (\citealp{Thall2008}).   

Several authors have considered both objective response and overall survival  as co-primary endpoints in cancer trials. \citet{Lai2015} proposed a single-arm phase II trial design with tumor response rate and a time-to-event outcome, such as overall survival or progression free survival. In their design, the dependence between the probability of response and the time-to-event outcome is modeled through a Gaussian copula.
\citet{Lai2012} proposed a  two-step sequential design in which  the response rate and the  time to the event are jointly modeled.
Their approach relates the response rate and the  time to the event  by means of a mixture model build on the basis of the Cox proportional hazards model assumption.  
\citet{Chen2020} presented a joint model for binary marker responses and survival outcomes for clustered data. 
They based the statistical inference on a multivariate penalized likelihood method and estimate the standard errors using a jackknife resampling method.

An additional challenge in immunotherapy trials lies in the fact that delayed effects are likely to be found, bringing the need of  
alternative methods accounting for the non-proportionality of the hazards (\citealp{Mick2015}). 
Statistics  that look at differences between integrated weighted survival curves, such as those defined by \citet{Pepe1989,Pepe1991} and extended by \citet{Gu1999}, are better suited to detect early or late survival differences and do not depend on the proportional hazards assumption.
In this paper, we aim to propose a class of two-sample statistics that  could be used in seamless phase II/III design  to jointly  evaluate the efficacy on binary and survival endpoints, 
even in the presence of delayed treatment effects.


The problem of how to analyze multiple outcomes has been widely discussed in the literature (\citealp{Dmitrienko2013,Alosh2014}). The classical approach is to restrict the attention to multiple testing procedures that control the probability of one or more false rejections, the so-called familywise error rate, which  guarantee the nominal significance level (\citealp{LehmannBook}). 
However, classical multiple testing procedures based on correcting the significance level  (e.g, Bonferroni procedure (\citealp{Bland1995})) may  not be appropriate since they 
do not take into account the potential association between the binary and survival outcomes and might lead to conservative designs.

Other alternative approaches have been developed allowing for the joint distribution of test statistics.
\citet{OBrien1984}  and \citet{Pocock1987} 
proposed global test statistics through the sum of individual statistics. 
\citet{OBrien1984}  developed a generalized least squares method by combining multiple statistics into a single hypothesis test when variables are normally distributed;
whereas  
\citet{Pocock1987} extended O'Brien's work to  asymptotically normal  test statistics.  
\citet{Hothorn2008} and \citet{Pipper2012} approached the problem of testing multiple hypothesis using parametric and semi-parametric models. 
\citet{Hothorn2008} used the limiting distribution of the parameter estimators to build upon the corresponding test statistics and their joint distribution. Based on that, their  approach corrects the significance level  by means of the simultaneous asymptotic normality of the test statistics. 
\citet{Pipper2012} proposed a procedure for evaluating the efficacy in trials with multiple endpoints of different types. Their procedure is based on simultaneous asymptotic normality of effect estimators from the  single-models for each endpoint together with multiple testing adjustments.

Extensive research has been done on joint modeling of longitudinal measurements and survival data (comprehensive overviews can be found in \citet{Tsiatis2004,Rizopoulos2012} and \citet{Papageorgiou2019}).
In most cases, the primary focus is on characterizing the association between the longitudinal and event time processes. The common framework is to relate the time-to-event and longitudinal outcomes through the proportional hazard model. 
Nevertheless, the relationship between binary response at a specific time point and survival outcome has received less attention (\citet{Chen2020}).


In this paper, we have followed the idea launched by \citet{Pocock1987} of combining multiple test statistics into a single hypothesis test. Specifically, we propose a class of statistics based on a weighted sum of a difference in proportions test and a weighted Kaplan-Meier test-based for the difference of survival functions. Our proposal adds versatility into the study design by enabling different follow-up periods for each endpoint, and flexibility by incorporating  weights. We define these weights to specify unequal priorities to the different endpoints and to anticipate the type of  time-to-event difference to be detected. 

This article is organized as follows. In Section 2 we  present the class of statistics for binary and time-to-event outcomes. In Section 3 we set out the assumptions and present the large sample distribution theory for the proposed statistics. In Section 4 we introduce different weights and discuss about their  choice.
We give an overview of our R package   \texttt{SurvBin} in Section 5 and 
illustrate our proposal with a recent immunotherapy trial in Section 6. 
In Section 7 we evaluate the performance of these statistics in terms of the significance level with a simulation study.
We conclude with a discussion. 

All the required functions to use these statistics have been implemented in R and have been made available at: https://github.com/MartaBofillRoig/SurvBin.

\section{A general class of binary and survival test statistics} \label{Sect2}

Consider a study comparing two groups, control group ($i=0$) and intervention group ($i=1$), 
each composed of $n^{(i)}$ individuals, and denote by $n=n^{(0)}+n^{(1)}$ the total sample size. Suppose that both groups are followed over the time interval  $[0,\tau]$ and are 
compared on the basis of the following two endpoints: the occurrence of an event $\varepsilon_b$ before $\tau_b$ ($0<\tau_b\leq \tau$), and the time to a different event $\varepsilon_s$ within the interval $[\tau_0,\tau]$ ($0\leq \tau_0< \tau$).
For  the $i$-th group ($i=0,1$), let   $ p^{(i)}\left( \tau_b\right)  $  be  the     probability of having   the event $\varepsilon_b$ before $\tau_b$, and  $S^{(i)}(\cdot)$ be the survival function of the time to the event  $\varepsilon_s$.

We consider the problem of testing  simultaneously $\mathrm{H}_{b,0}$:   $p^{(0)}(\tau_b)=p^{(1)}(\tau_b) $ and $\mathrm{H}_{s,0}$: $S^{(0)}(t) =S^{(1)}(t), \forall t \in [\tau_0,\tau] $,
aiming to demonstrate either a higher probability of the occurrence of $\varepsilon_b$ or an improved survival with respect to $\varepsilon_s$ in the intervention group. The hypothesis problem  can then be formalized as: 
\begin{equation} \label{HT}
	\begin{cases}
		\mathrm{H}_0: & p^{(0)}(\tau_b)=p^{(1)}(\tau_b)  \ \text{ and } \  S^{(0)}(t) =S^{(1)}(t), \ \forall t \in [\tau_0,\tau]  \\
		\mathrm{H}_1: & p^{(0)}(\tau_b) < p^{(1)}(\tau_b)   \ \text{ or } \  S^{(0)}(t)   \leq S^{(1)}(t), \ \ \forall t \in [\tau_0,\tau], \\
		& \hspace{35mm} \exists t \in [\tau_0,\tau], \ S^{(0)}(\cdot) < S^{(1)}(\cdot)
	\end{cases} 
\end{equation} 

We propose a class  of statistics --hereafter called $\mathcal{L}$-class-- as a weighted linear combination of the difference of proportions statistic for the binary outcome
and the integrated weighted difference of two survival functions for the time-to-event outcome, 
as follows, 
\begin{equation} \label{Lclass} 
	\mathrm{\textbf{U}}_{n}^{\omega}(\tau_0,\tau_b,\tau; \hat{Q})
	=  {\omega}_b \cdot \frac{U_{b,n}\left(\tau_b\right)}{\hat{\sigma}_b}  
	+  {\omega}_s \cdot \frac{U_{s,n}(\tau_0,\tau;\hat{Q})}{\hat{\sigma}_s} 
\end{equation}
for some real numbers $\omega_b,\omega_s\in (0,1)$, such that  $\omega_b+\omega_s=1$, and where: 
\begin{eqnarray} \label{bintest}
	U_{b,n}\left(\tau_b\right) &=&   \sqrt{\frac{n^{(0)}n^{(1)}}{n}} \left(\hat{p}^{(1)}\left(\tau_b\right)  -  \hat{p}^{(0)}\left(\tau_b\right)\right) \\ \label{survtest}
	U_{s,n}(\tau_0,\tau;\hat{Q})   &=&   
	\sqrt{\frac{n^{(0)}n^{(1)}}{n}}\left( \int_{\tau_0}^{\tau} \hat{Q}(t) \cdot\left( \hat{S}^{(1)}(t)-\hat{S}^{(0)}(t)\right)  dt \right) 
\end{eqnarray} 
denoting by $\hat{p}^{(i)}(\tau_b)$ the estimated proportion of events $\varepsilon_b$ before  $\tau_b$, and by $\hat{S}^{(i)}(\cdot)$ the  Kaplan-Meier estimator of $S^{(i)}(\cdot)$ for group $i$. 
The estimates $\hat{\sigma}_b^2$ and $\hat{\sigma}_s^2$ are such that 
converge in probability to $\sigma_b^2$ and  $\sigma_s^2$, respectively,
as $n\rightarrow +\infty$, where $\sigma_b^2$ and $\sigma_s^2$ represent the variances of $U_{b,n}\left(\tau_b\right)$ and $U_{s,n}(\tau_0,\tau;\hat{Q})$, respectively. Both theoretical and estimated expressions for the variances of $U_{b,n}\left(\tau_b\right)$ and $U_{s,n}(\tau_0,\tau;\hat{Q})$ will be given in Section \ref{Sect3.Asymptotics} (see equations (\ref{sigb},~\ref{sigs}) for the theoretical expressions and (\ref{sigb_est},~\ref{sigs_est}) for the estimates).
The term $\hat{Q}(\cdot)$  is a possibly random function which converges pointwise in probability to a deterministic function $Q(\cdot)$.
For ease of notation, and letting $\omega=\left(\omega_b,\omega_s\right)$,  we will suppress the dependence on $\tau_0,\tau_b,\tau$ and  use instead $\mathrm{\textbf{U}}_{n}^{\omega}(\hat{Q})$, $U_{b,n}$, $U_{s,n}(\hat{Q})$. 
Note that $\hat{p}^{(i)}(\tau_b)$, $\hat{S}^{(i)}(\cdot)$, $\hat{\sigma}_b$ and $\hat{\sigma}_s$ depend on the sample size $n^{(i)}$, but it has been omitted in notation for short.

The weights $\omega$  control the relative relevance of each outcome -if any- and  the random weight function $\hat{Q}(\cdot)$ serves two purposes: to specify the type of survival differences that may exist between groups and to stabilize the variance of the difference of the two Kaplan-Meier functions. Some well-known special cases of $\hat{Q}(\cdot)$ are:
\begin{enumerate}[(i)]
	\item $\hat{Q}(t)=\hat{G}(t-)$, 	where $\hat{G}(t-)$ is the pooled Kaplan-Meier estimator for the censoring distribution.
	This choice of $\hat{Q}(t)$ down-weights the contributions on those times where the  censoring  is heavy.  
	\item $\hat{Q}(t)=\hat{S}(t-)^\rho \cdot (1-\hat{S}(t-))^\gamma$,  where $\rho,\gamma\geq0$ and $\hat{S}(t-)$ is the pooled Kaplan-Meier estimator for the survival function. This $\hat{Q}(t)$  corresponds to
	the weights of the $G^{p,q}$ family
	(\citealp{Fleming1991}). Then, for instance, if $\rho=1$ and $\gamma=0$, $\hat{Q}(t)$ emphasizes early differences between survival functions; whereas late differences could be highlighted with $\rho=0$ and $\gamma=1$. 
	\item $\hat{Q}(t)=\bar{Y}(t-) $,  where $\bar{Y}(t-)$ denotes the number of individuals at risk of $\varepsilon_s$ at time $t$. In this case $\hat{Q}(t)$ accentuates the information at the beginning of the survival curve allowing early failures to receive more weight than later failures.   
\end{enumerate} 
We state the precise conditions for the weight function $\hat{Q}(\cdot)$ in Section \ref{Sect3.Asymptotics} and postpone the discussion about the choice of $\hat{Q}(\cdot)$ and $\omega=(\omega_b,\omega_s)$ to Section \ref{Sect4.Weights}.

\begin{figure}[p!] 
	\centering
	\includegraphics[width=0.6\linewidth]{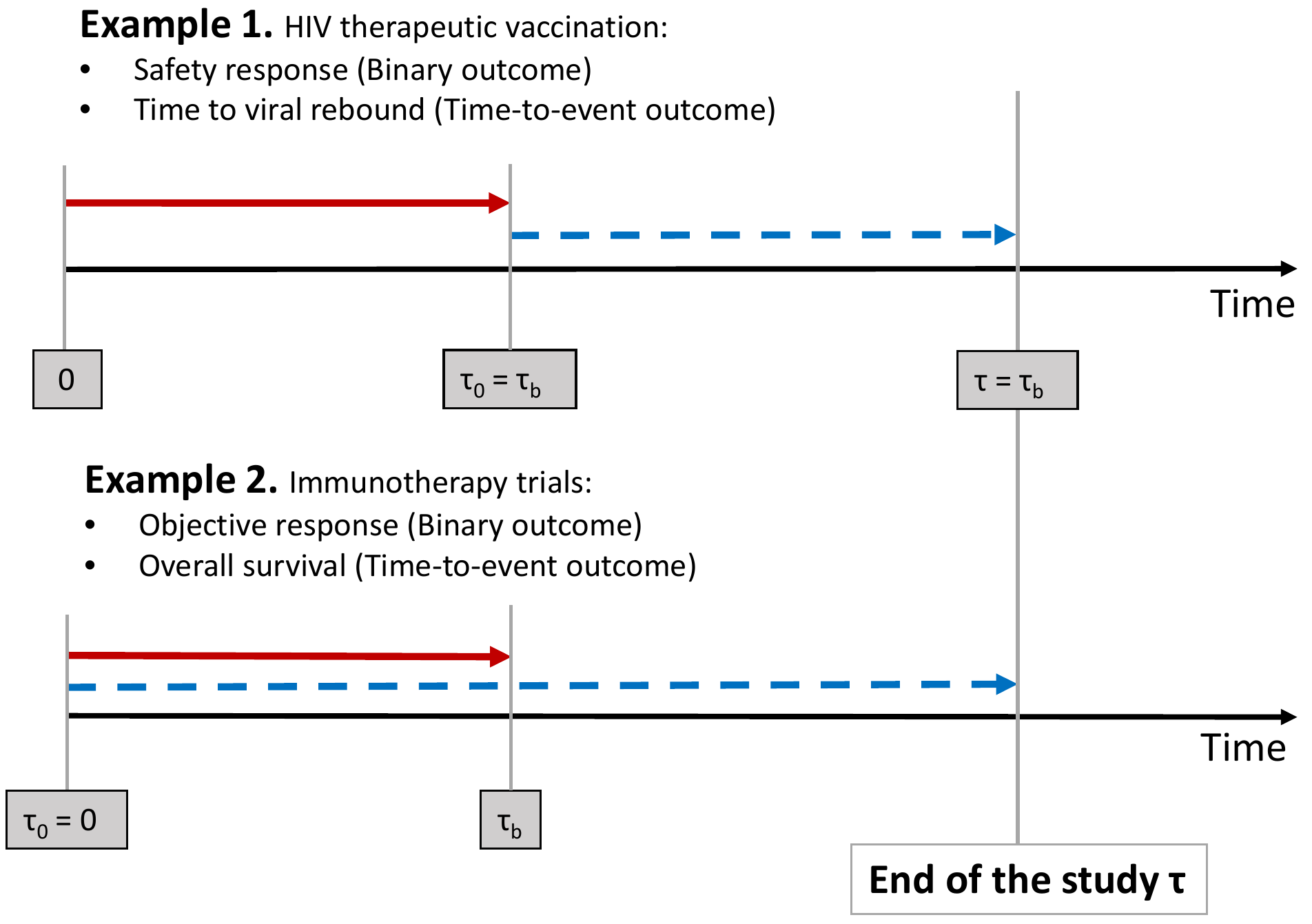} 
	\caption{Illustration of two different follow-up configurations, the red and blue arrows represent the time-frame for binary and time-to-event outcomes, respectively. The red line goes from the study starts (at time-point $0$) until the binary outcome is evaluated ($\tau_b$). The blue (dashed) line goes from the time-to-event information begins to be collected ($\tau_0$) to the end of the study ($\tau$).}
	\label{fig:examples}
\end{figure}

The statistics in the $\mathcal{L}$-class are defined for  different follow-up configurations based on different choices of: the overall follow-up period, $\tau$; the time where the binary event is evaluated, $\tau_b$; and the origin time for the survival outcome, $\tau_0$; taking into account that $0<\max\{\tau_0, \tau_b\}<\tau$. There are however no restrictions on whether or not these periods overlap
and, if they do, how much and when. We illustrate two different situations with different configurations for $\tau_0,\tau_b,\tau$ in Figure  \ref{fig:examples}.   
The first case is exemplified by an HIV therapeutic vaccination study where  safety-tolerability response (binary outcome) and  time-to-viral rebound (survival outcome) are outcomes of interest. Whereas the safety-tolerability is evaluated at week 6 ($\tau_b=6$), the time-to-viral rebound is evaluated from week 6 to 18 ($\tau_0=6$ and $\tau=18$) (\citealp{Jong2019}). The second example in the area of immunotherapy trials includes a binary outcome (objective response), evaluated at month 6, and overall survival,  evaluated from randomization until year 4 ($\tau_0=0$, $\tau_b=0.5$ and $\tau=4$)  (\citealp{Hodi2010}).   

The $\mathcal{L}$-class statistics includes several  statistical tests. 
If $\tau_0=0$, $\tau_b=\tau$ and $\omega_b=\omega_s$, then, $\mathrm{\textbf{U}}_{n}^{\omega}(\hat{Q})$ corresponds to the global test statistic proposed by \citet{Pocock1987}. If $\varepsilon_b=\varepsilon_s$, $\tau_0=\tau_b$, and $\omega_b=\omega_s$, the statistic $\mathrm{\textbf{U}}_{n}^{\omega}(\hat{Q})$ is the equivalent of the linear combination test of \citet{Logan2008} when there is no censorship until $\tau_b$ for testing for differences in survival curves after a pre-specified time-point.

\section{Large sample results} \label{Sect3.Asymptotics} 

In this section,
we derive the asymptotic distribution of the $\mathcal{L}$-class of statistics given in \eqref{Lclass} under the null hypothesis and under contiguous alternatives, present an estimator of their asymptotic variance, and discuss the consistency of the $\mathcal{L}$-statistics against any alternative hypothesis of the form of $\mathrm{H}_1$ in \eqref{HT}.  
We start the section with the conditions we require for the $\mathcal{L}$-class of statistics. 
In order to make the paper more concise and more readable, 
proofs and technical details are in the Supplementary material.

\subsection{Further notation and Assumptions} \label{Sect3.Assumptions} 

We consider two independent random  samples of $n^{(i)}$ ($i=0,1$) individuals and for each we denote the binary response by  $X_{ij}=I \{ \varepsilon_b$ has occurred$\}$,  the time to $\varepsilon_s$ by $T_{ij}$  and the censoring time by $C_{ij}$ for $j=1,..., n^{(i)}$ and  where  $I \{\cdot\}$ is  the usual 0/1 indicator function.
Assuming that $T_{ij}$  is non-informatively right-censored by $C_{ij}$,  the observable data are  summarized by $\{X_{ij}, T_{ij} \wedge C_{ij}, \delta_{ij}\}$,  where $\delta_{ij} = I \{T_{ij} \wedge C_{ij}=T_{ij}\}$ and $a\wedge b = \min(a,b)$.
Suppose as well  that $X_{ij}$ is independent of  $C_{ij}$ and that the occurrence of the survival and censoring times, $T_{ij}$ and $C_{ij}$, does not prevent to assess the binary response, $X_{ij}$.

Denote by $G^{(i)}(\cdot)$ and $\hat{G}^{(i)}(\cdot)$  the censoring survival function  and the Kaplan-Meier estimator for the censoring times, respectively. 
As we will see in the next section, the distribution of the $\mathcal{L}$-statistics relies, among others, on the survival function for  those patients who respond to the binary endpoint. We then introduce here the survival function for responders as  $S_X^{(i)}(t)=$ P$(T_{ij}>t|X_{ij}=1)$ $(t>\tau_b)$.  

Furthermore we assume that: (i) $S^{(i)}(\tau)>0$, $S_X^{(i)}(\tau)>0$ and $G^{(i)}(\tau)>0$;  (ii) the limiting fraction of the total sample size is non-negligible, i.e., ${\displaystyle \lim_{n \rightarrow + \infty}} n^{(i)}/n = \pi^{(i)} \in (0,1)$; and (iii) $Q(\cdot)$ is a nonnegative piecewise continuous with finitely discontinuity points. For all the continuity points in $[0,\tau]$, $\hat{Q}(t)$ converges in probability to $Q(t)$  as $n\rightarrow + \infty$. Moreover,  $\hat{Q}(\cdot)$ and $Q(\cdot)$ are functions of total variation bounded in probability.

Finally, we introduce the counting process $\bar{N}^{(i)}(t) = \sum_{j=1}^{n^{(i)}}N_{ij}(t) = \sum_{j=1}^{n^{(i)}} I \{T_{ij} \wedge C_{ij}\leq t, \delta_{ij}=1 \}$ as the number of observed events that have occurred by time $t$ for the $i$-th group ($i=0,1$)  and $\bar{Y}^{(i)}(t) = \sum_{j=1}^{n^{(i)}}Y_{ij}(t) = \sum_{j=1}^{n^{(i)}} I \{T_{ij} \wedge C_{ij}\geq t  \}$ as the number of subjects at risk at time $t$ for the $i$-th group.  We define ${y}^{(i)}(s)=\mathrm{E}\left(Y_{ij}(s)\right) $ and suppose that ${y}^{(i)}(\tau)>0$.

Remark: Throughout the paper and to refer to the group $i$ ($i=0,1$), we will use subindexes for the individual observations and stochastic processes, as in $X_{ij}$, while we will use  superindexes in parentheses for the functions and parameters, as in $S^{(i)}(\cdot)$.


\subsection{Asymptotic distribution}

In order to derive the asymptotic distribution of the statistic $\mathrm{\textbf{U}}_{n}^{\omega}(\hat{Q})$, 
we first note that $\mathrm{\textbf{U}}_{n}^{\omega}(\hat{Q})$ can be approximated by $\mathrm{\textbf{U}}_{n}^{\omega}({Q})$,  the same statistic with the weights replaced by its deterministic function. 

\begin{lemma} \label{lemma}
	Let $\mathrm{\textbf{U}}_{n}^{\omega}(Q)$ be the statistic defined by:	
	\begin{equation} \label{stapprox} 
		\mathrm{\textbf{U}}_{n}^{\omega}(Q)
		=  {\omega}_b \cdot \frac{U_{b,n}}{\hat{\sigma}_b}  
		+  {\omega}_s \cdot \frac{U_{s,n}(Q)}{\hat{\sigma}_s} 
	\end{equation} 
	where $U_{b,n}$ is the statistic given in \eqref{bintest} and 	$U_{s,n}(Q) $ is the statistic given in \eqref{survtest} with  $\hat{Q}(t)$ replaced by  $Q(t)$, that is:
	\begin{eqnarray} \label{Usk}
		U_{s,n}(Q) &=& \sqrt{\frac{n^{(0)}n^{(1)}}{n}}\left( \int_{\tau_0}^{\tau} Q(t) \cdot\left( \hat{S}^{(1)}(t)-\hat{S}^{(0)}(t)\right)  dt \right) 
	\end{eqnarray} 
	for some real numbers $\omega_b,\omega_s\in (0,1)$, such that  $\omega_b+\omega_s=1$, and for a function 
	$Q(\cdot)$ satisfying the conditions outlined in Section \ref{Sect3.Assumptions}. 	
	Then, the $\mathcal{L}$-statistic $\mathrm{\textbf{U}}_{n}^{\omega}(\hat{Q})$, given in \eqref{Lclass}, can be written as:
	$$\mathrm{\textbf{U}}_{n}^{\omega}(\hat{Q}) = \mathrm{\textbf{U}}_{n}^{\omega}(Q) + \omega_s\cdot \frac{E_n}{\hat{\sigma}_s}$$
	where 
	$$E_n = \sqrt{\frac{n^{(0)}n^{(1)}}{n}} 
	\int_{\tau_0}^{\tau} ( \hat{Q}(t)-Q(t))  \cdot (\hat{S}^{(1)}(t)-\hat{S}^{(0)}(t))  dt $$ 
	converges in probability to $0$.
	Hence, the asymptotic distribution of the statistic  $\mathrm{\textbf{U}}_{n}^{\omega}(Q)$ is the same as that of $\mathrm{\textbf{U}}_{n}^{\omega}(\hat{Q})$. 
\end{lemma}  

Roughly speaking, thanks to this theorem we can ignore the randomness of $\hat{Q}(\cdot)$ and use $\mathrm{\textbf{U}}_{n}^{\omega}(Q)$ to obtain the limiting distribution of $\mathrm{\textbf{U}}_{n}^{\omega}(\hat{Q})$.  
In what follows,
we state the asymptotic distributions under the null hypothesis in Theorem \ref{th-NullD} and  under a sequence of contiguous alternatives in Theorem \ref{th-ContD}.
\begin{theorem} \label{th-NullD}
	Let $\mathrm{\textbf{U}}_{n}^{\omega}(\hat{Q})$ be the statistic defined in \eqref{Lclass}. 
	Under the conditions outlined in \ref{Sect3.Assumptions}, if the null hypothesis $ \mathrm{H}_{0}:  \mathrm{H}_{s,0} \cap \mathrm{H}_{b,0}$ holds, $\mathrm{\textbf{U}}_{n}^{\omega}(\hat{Q})$ converges in distribution, as $n\rightarrow + \infty$,  to a normal distribution as follows:
	\begin{eqnarray*}
		\mathrm{\textbf{U}}_{n}^{\omega}(\hat{Q}) 
		\rightarrow    
		N \left(  0, \omega_b^2 + \omega_s^2 +2\omega_b\omega_s \cdot\frac{\sigma_{bs} }{\sigma_{b} \cdot \sigma_{s} }  \right) 
	\end{eqnarray*}
	where  $\sigma_b^2 $, $\sigma_{s}^2 $ stand for the variances of $U_{b,n}$ and $U_{s,n}(Q)$, respectively, and $\sigma_{bs} $ is the covariance between $U_{b,n} $ and $U_{s,n}(Q)$. Their corresponding expressions are given by:
	\begin{eqnarray}
		\sigma_b^2  &=&   \sum_{i=0,1}\big(1-\pi^{(i)} \big)   p^{(i)}\left(\tau_b\right)\left(1-p^{(i)}\left(\tau_b\right) \right)  
		\label{sigb} \\  
		\sigma_{s}^2   &=&    - \sum_{i=0,1}\big(1-\pi^{(i)} \big)   \int_{\tau_0}^{\tau} \frac{(K_\tau^{(i)}(t))^2 }{(S^{(i)}(t))^2 G^{(i)}(t)}  dS^{(i)}(t)  
		\label{sigs}\\
		\sigma_{bs}   &=&   \sum_{i=0,1} \big(1-\pi^{(i)} \big) \cdot \Big( I\{ \tau_{\max} = \tau_b \} \cdot 
		\int_{\tau_0}^{\tau_b}   \frac{K_{\tau_b}^{(i)}(t)}{S^{(i)}(t)} \cdot \left(p_N^{(i)}(t)-p^{(i)}\left(\tau_b\right)  \right) dS^{(i)}(t)  
		\Big. 
		\nonumber\\
		&&   \left. 
		+ \int_{\tau_{\max}}^{\tau}
		\frac{ K_\tau^{(i)}(t) }{S^{(i)}(t) } \cdot p^{(i)}\left(\tau_b\right) \left( dS_X^{(i)}(t) -dS^{(i)}(t)\right) \right) 
		\label{sigbs}  
	\end{eqnarray} 
	where 
	$\tau_{\max} = \max(\tau_0,\tau_b)$, $K_{\tau_*}^{(i)}(t)=\int_{t}^{\tau_*} Q(u) S^{(i)}(u)du$ 
	($\tau_*=\tau$ or $\tau_b$),  \\
	$p_N^{(i)}(t)=P(X_{ij}=1| dN_{ij}(t)=1)$, and $S_X^{(i)}(t) =P(T_{ij}>t|X_{ij}=1)$ for $i=0,1$. 
\end{theorem}

Recall that $\sigma_b^2$, $\sigma_{s}^2$, and $\sigma_{bs}$ depend on $\tau_0,\tau_b,\tau$, but we omit them for notational simplicity.

\begin{theorem} \label{th-ContD}
	Let $\mathrm{\textbf{U}}_{n}^{\omega}(\hat{Q})$ be the statistic defined in \eqref{Lclass}. 
	Under the conditions outlined in 3.1., consider the following sequences of contiguous alternatives for both binary and time-to-event hypotheses  satisfying,	as $n \rightarrow + \infty$: 
	$$\sqrt{n}(p_n^{(1)}-p^{(0)}) \rightarrow g$$
	and 
	$$\sqrt{n}( S_{n}^{(1)}(t)-S^{(0)}(t)) \rightarrow \mathcal{G}(t)$$
	for some constant $g \in \mathbb{R}^+$ and bounded function $\mathcal{G}(\cdot) \in \mathbb{R}^+$, and $\forall t \in [\tau_0,\tau]$.
	Then,  under contiguous alternatives of the form:
	\begin{eqnarray}
		\mathrm{H}_{1,n}: \sqrt{n}(p_n^{(1)}-p^{(0)})= g &   \text{ and }   &   \sqrt{n}( S_{n}^{(1)}(t)-S^{(0)}(t))= \mathcal{G}(t), \ \ \ \forall t \in [\tau_0,\tau]
	\end{eqnarray} 
	we have that:  
	$$\mathrm{\textbf{U}}_{n}^{\omega}(\hat{Q}) \rightarrow N   \left(   \omega_b g + \omega_s \int_{\tau_0}^{\tau}   Q(t) \mathcal{G}(t) dt,  \omega_b^2 + \omega_s^2 +2\omega_b\omega_s\frac{\sigma_{bs}}{\sigma_{b} \cdot \sigma_{s}}  \right)$$ 
	in distribution as $n\rightarrow + \infty$, where
	$\sigma_{b}^2$, $\sigma_{s}^2$ and $\sigma_{bs}$ are given in \eqref{sigb}, \eqref{sigs} and \eqref{sigbs}, respectively. 
\end{theorem}

The covariance in \eqref{sigbs} involves the conditional probabilities $S_X^{(i)}(t)$  and $p_N^{(i)}(t)$, while $S_X^{(i)}(t)$ represents the    survival function for responders --patients that have had the binary event $\varepsilon_b$--, $p_N^{(i)}(t)$ stands for the probability of being a responder among patients experiencing $\varepsilon_s$ at $t$.  
Also note that, if $\tau_b<\tau_0$, 
the survival experience starts after the binary event has been evaluated and only involves the second integral in \eqref{sigbs}.

We notice that the efficiency of the $\mathcal{L}$-statistics, $\mathrm{\textbf{U}}_{n}^{\omega}(\hat{Q})$,  under contiguous alternatives is driven by the non-centrality parameter $ \mu_{c} = \omega_b g + \omega_s \int_{\tau_0}^{\tau}  Q(t) \mathcal{G}(t) dt$, that is,  by the sum of the weighted non-centrality parameters of $U_{b,n}$  and $U_{s,n}(\hat{Q})$.

\subsection{Variance estimation and consistency}

We now describe how to use the $\mathcal{L}$-statistics  to test $\mathrm{H}_0$  versus $\mathrm{H}_1$ given in \eqref{HT}.  
Theorem \ref{th-estimation} gives a consistent estimator of the asymptotic variance of $\mathrm{\textbf{U}}_{n}^{\omega}(\hat{Q})$, and  Theorem \ref{th-consistency}  presents the standardized $\mathcal{L}$-statistics to test $\mathrm{H}_{0}:  \mathrm{H}_{s,0} \cap \mathrm{H}_{b,0}$.

\begin{theorem} \label{th-estimation}
	Let $\mathrm{\textbf{U}}_{n}^{\omega}(\hat{Q})$ be the statistic defined in \eqref{Lclass}, and let $\sigma_b^2$, $\sigma_{s}^2$ and $\sigma_{bs}$ be the variances and covariance given in \eqref{sigb}, \eqref{sigs} and \eqref{sigbs}, respectively. The  asymptotic variance of $\mathrm{\textbf{U}}_{n}^{\omega}(\hat{Q})$, given in Theorem \ref{th-NullD}, can be consistently estimated by:
	\begin{eqnarray} \label{varLclass}
		\widehat{\mathrm{Var}}(\mathrm{\textbf{U}}_{n}^{\omega}(\hat{Q})) &=& \omega_b^2 + \omega_s^2 +2\omega_b\omega_s\frac{\hat{\sigma}_{bs} }{\hat{\sigma}_{b} \cdot \hat{\sigma}_{s} }
	\end{eqnarray}
	where $\hat{\sigma}_{b}$, $\hat{\sigma}_{s}$, and $\hat{\sigma}_{bs}$ denote the estimates of $\sigma_{b}$, $\sigma_{s}$  and $\sigma_{bs}$, and are given by:
	\begin{eqnarray}  
		\hat{\sigma}_{b}^2 &=&   \hat{p}\left(\tau_b\right)\left(1-\hat{p}\left(\tau_b\right) \right)  
		\label{sigb_est}  \\ 
		\hat{\sigma}_{s}^2 
		&=&  
		-  \int_{\tau_0}^{\tau}   \frac{(\hat{K}_\tau(t))^2 }{\hat{S}(t)\hat{S}(t-)} \cdot \frac{n^{(0)}\hat{G}^{(0)}(t-)+n^{(1)}\hat{G}^{(1)}(t-)}{\hat{G}^{(0)}(t-)\hat{G}^{(1)}(t-)}  d\hat{S}(t) 
		\label{sigs_est} \\
		\hat{\sigma}_{bs}    &=&  
		-  \int_{\tau_0}^{\tau_b}  \hat{K}_{\tau_b}(t) 
		\left( 
		\sum_{i=0,1}  \frac{n-n^{(i)}}{n} \cdot \hat{\lambda}_{X,T}^{(i)}(t)dt+ \frac{\hat{p}\left(\tau_b\right) \cdot d\hat{S}(t) }{\hat{S}(t)} \right)  
		\nonumber\\
		&&  
		+  \int_{\tau_b}^{\tau}  
		\frac{\hat{K}_{\tau}(t) \cdot\hat{p}\left(\tau_b\right)}{\hat{S}(t-) }  \left( - \frac{\hat{S}(t-) \cdot d\hat{S}(t)}{\hat{S}(t)} 
		+ \sum_{i=0,1} \frac{n-n^{(i)}}{n} \cdot \frac{\hat{S}_X^{(i)}(t-)\cdot d\hat{S}_X^{(i)}(t)}{\hat{S}_X^{(i)}(t)} 
		\right)  
		\label{sigbs_est}
	\end{eqnarray} 
	where $\hat{K}_{\tau_*}(t)=\int_{t}^{\tau_*} \hat{Q}(u) \hat{S}(u)du$  ($\tau_*=\tau$ or $\tau_b$), $\hat{S}_X^{(i)}(t)$ is the Kaplan-Meier estimator of $S_X^{(i)}(t)$; and 
	$\hat{\lambda}_{X,T}^{(i)}(t)$ is the estimator of 	$\lambda_{X,T}^{(i)}(t) = \lim_{dt\rightarrow0} P(X_{ij}=1, t\leq T_{ij} < t +dt |T_{ij}>t)/dt$.  Kernel-density methods are used in the estimation of  ${\lambda}_{X,T}^{(i)}(t)$. 
\end{theorem} 

\begin{theorem} \label{th-consistency}
	Let $\mathrm{\textbf{U}}_{n}^{\omega}(\hat{Q})$ be the statistic defined in \eqref{Lclass},
	and let $\widehat{\mathrm{Var}}(\mathrm{\textbf{U}}_{n}^{\omega}(\hat{Q}))$ be the variance estimator given in \eqref{varLclass}.
	Consider the global null hypothesis $\mathrm{H}_{0}$ (\eqref{HT}) and let the normalized statistic of $\mathrm{\textbf{U}}_{n}^{\omega}(\hat{Q})$ be:
	\begin{equation} \label{Lclass.std}
		\mathrm{\textbf{U}}_{n}^{\omega}(\hat{Q})\bigg/\sqrt{\widehat{\mathrm{Var}}(\mathrm{\textbf{U}}_{n}^{\omega}(\hat{Q}))} 
	\end{equation}  
	Then, the statistic defined in \eqref{Lclass.std} converges in distribution to a standard normal distribution. 
	Moreover, for positive $Q(\cdot)$, the statistic is consistent against any alternative hypothesis of the form of $\mathrm{H}_1$ in \eqref{HT} which contemplate differences and stochastic ordering alternatives for the binary and time-to-event outcomes, respectively. 
\end{theorem}

We are presenting here a pooled variance estimator of $\mathrm{\textbf{U}}_{n}^{\omega}(\hat{Q})$.   An unpooled variance estimator is proposed
in the Supplementary Material.

Theorem \ref{th-consistency} can be used to test
the global null hypothesis   $\mathrm{H}_{0}:  \mathrm{H}_{s,0} \cap \mathrm{H}_{b,0}$ by comparing $\mathrm{\textbf{U}}_{n}^{\omega}(\hat{Q})\big/\sqrt{\widehat{\mathrm{Var}}(\mathrm{\textbf{U}}_{n}^{\omega}(\hat{Q}))}$ to a standard normal distribution.


\section{On the choice of weights} \label{Sect4.Weights}

An important consideration when applying the statistics proposed in this paper is the choice of the weight functions. The $\mathcal{L}$-class of statistics involves the already mentioned random weight function $ \hat{Q}(t)$ and deterministic weights $\omega=(\omega_b,\omega_s)$.
These weights are defined according to different purposes and have different roles  into the statistic $\mathrm{\textbf{U}}_{n}^{\omega}(Q)$. In this section, we include different weights and discuss some of their strengths as well as shortcomings.  The list provided is not exhaustive, other weights are possible and might be useful in specific circumstances.

\subsection{Choice of $\omega=(\omega_b,\omega_s)$}

The purpose of the weights  $\omega$ is to prioritize the binary and the  time-to-event outcomes. They have to be specified in advance according to the research questions. Whenever the two outcomes are equally relevant, we should choose $\omega_b=\omega_s=0.5$. In this case the statistics will be optimal whenever the standardized effects on both outcomes coincide.

\subsection{Choice of $\hat{Q}(\cdot)$}

The choice of $\hat{Q}(\cdot)$ might be very general as long as $\hat{Q}(\cdot)$ converges  in probability to a function $Q(\cdot)$, and both $\hat{Q}(\cdot)$ and $Q(\cdot)$ satisfy the conditions outlined in \ref{Sect3.Assumptions}. In this section, we center our attention on a family of $\hat{Q}(\cdot)$ weights of the form: 
$$\hat{Q}(t)= \hat{f}(t) \cdot \hat{v}(t),$$
where: (i) $\hat{f}(\cdot)$ is a data-dependent function that converges, in probability to $f(\cdot)$, a nonnegative piecewise continuous function with bounded variation on $[0,1]$.  The term $\hat{f}(t)$ takes care of the expected differences between survival functions  and can be used as well to emphasize  some parts of the follow-up according to the time-points ($\tau_0,\tau_b,\tau_s$); 
(ii) the weights $\hat{v}(\cdot)$  converge  in probability to a deterministic positive bounded weight function  $v(\cdot)$.  
The main purpose of the weight $\hat{v}(t)$ is to ensure the stability of the variance of the difference of the two Kaplan-Meier functions. To do so,  we make the additional assumption that: 
$$ 
|v(t)| \leq \Gamma \cdot G^{(i)}(t)^{1/2 + \delta}   \ \ \ \text{ and } \ \ \
|\hat{v}(t)| \leq \Gamma \cdot \hat{G}^{(i)}(t)^{1/2 + \delta}
$$
for all $t \in [\tau_0,\tau] $, $i=0,1$ and for some constants $\Gamma,\delta>0$.

Different choices of $\hat{f}(t) $ yield other known statistics. For instance, if $f(\cdot)=1$, $U_{s,n}(\hat{Q})$ corresponds to the Weighted Kaplan-Meier statistics (\citealp{Pepe1989,Pepe1991}). Whenever $\hat{f}$ and  $\hat{v}$ correspond to the weights \eqref{f.w} and \eqref{w.w}, respectively, introduced below, we have the statistic proposed by \citet{Shen2001}.  Furthermore, note that  the weight functions of the form $\hat{Q}(t)= \hat{f}(t) \cdot \hat{v}(t)$
are similar to those proposed by \citet{Shen2001}; while they assume that $\hat{f}$ is a bounded continuous function,  
we assume that $\hat{f}(\cdot)$ is a nonnegative piecewise continuous function with bounded variation on $[0,1]$, and instead of only considering the Pepe and Fleming weight function corresponding to \eqref{f.w}, we also allow for different weight functions $\hat{v}(t)$. Finally, if the random quantity $\hat{Q}(t)$ is omitted $U_{s,n}(\hat{Q})$ corresponds to
the difference of restricted mean survival times from $\tau_0$ to $\tau$.

In what follows, we outline different choices of  $\hat{v}(t)$ and  $\hat{f}(t)$, together with a brief discussion for each one:
\begin{itemize}
	\item We require $\hat{v}(t)$ to be small towards the end of the observation period if censoring is heavy. 
	The usual weight functions $\hat{v}(t)$ involve Kaplan-Meier estimators of the censoring survival functions.  
	The most common weight functions are:
	\begin{eqnarray}\label{w.w}
		\hat{v}_{c}(t) = \frac{n\hat{G}^{(0)}(t-)\hat{G}^{(1)}(t-)}{n^{(0)}\hat{G}^{(0)}(t-)+n^{(1)}\hat{G}^{(1)}(t-)}
	\end{eqnarray} 
	and $\hat{v}_{\sqrt{}}(t) = \sqrt{\hat{v}_{c}(t)}$, both proposed by Pepe and Fleming.
	Among other properties, $\hat{v}_{c}(\cdot)$ has been proved to be a competitor to the logrank test for the proportional hazards alternative (\citealp{Pepe1989}). 
	Note that if the censoring survival functions are equal for both groups and the sampling design is balanced ($n^{(0)}=n^{(1)}$), then, the differences in Kaplan-Meier estimators are weighted by the censoring survival function, that is, ${w}(t) ={C}(t)={C}^{(i)}(t)$ for $i=0,1$. Also note that ${w}(t)= 1$ for uncensored data.
	\item 
	Analogously to \citet{Fleming1991} statistics, $\hat{f}(t)$ could be used to specify the type of expected differences between survival functions. That is, if we set: 
	\begin{eqnarray}\label{f.w}
		f(\hat{S}(t-)) = \hat{S}(t-)^\rho (1-\hat{S}(t-))^\gamma,
		\ \ \ \rho, \gamma \geq 0
	\end{eqnarray}
	the choice
	$\rho> 0$, $\gamma = 0$
	leads to a test to detect early differences, while
	$\rho= 0$, $\gamma > 0$
	leads to a test to detect late differences; and
	$\rho=\gamma=0$ leads to a test evenly distributed over time and corresponds to the weight function of the logrank. 
	\item In order to put more emphasis on those times after the binary follow-up period we might consider: 
	$$
	f(t)=
	\begin{cases}
	a,  & t < \tau_b\\
	1-a,  & t  \geq \tau_b
	\end{cases}
	$$ 
	for $a<0.5$.
\end{itemize}


\section{Implementation} 

We have developed the \texttt{SurvBin} package to facilitate the use of the $\mathcal{L}$-statistics and is now available on GitHub.
The \texttt{SurvBin} package contains three key functions: 
\texttt{lstats} to compute the standardized $\mathcal{L}$-statistic, {\small$\mathrm{\textbf{U}}_{n}^{\omega}(\hat{Q})\big/\sqrt{\widehat{\mathrm{Var}}(\mathrm{\textbf{U}}_{n}^{\omega}(\hat{Q}))}$}; and 
\texttt{bintest} and 
\texttt{survtest}  for the univariate binary and survival statistics \eqref{bintest} and \eqref{survtest}, {\small$U_{b,n}\left(\tau_b\right)\big/\hat{\sigma}_b$} and {\small$U_{s,n}(\tau_0,\tau;\hat{Q})\big/\hat{\sigma}_s$}, 
respectively.  
The \texttt{SurvBin} package also provides the functions \texttt{survbinCov}, that can be used to calculate  $\hat{\sigma}_{bs}$; and \texttt{simsurvbin} for simulating bivariate binary and survival data.

The main function \texttt{lstats}  can be called by:
\begin{center}
	\texttt{lstats(time, status, binary, treat, tau0, tau, taub, rho, gam, eta, wb, ws, var\_est)}
\end{center} 
where \texttt{time, status, binary} and \texttt{treat} are
vectors of the right-censored data, the status indicator, the binary data and the treatment group indicator, respectively;
\texttt{tau0, tau, taub} denote the follow-up configuration; 
\texttt{wb, ws} are the weights $\omega$; 
\texttt{rho, gam, eta} are scalar parameters that controls the weight $\hat{Q}(t)$ which is given by $\hat{Q}(t)=\hat{G}(t-)^\eta \cdot \hat{S}(t-)^\rho \cdot (1-\hat{S}(t-))^\gamma$;
and \texttt{var\_est} indicates the variance estimate to use (\texttt{pooled} or \texttt{unpooled}).

In this work, we estimate $\lambda_{X,T}^{(i)}(t)$
by means of the Epanechnikov kernel function, and the local bandwidth selection and the boundary correction described by \citet{Muller1994} 
by using the \texttt{muhaz} package (\citealp{Hess2019}). 


\section{Examples} \label{Section.Example} 

Melanoma has been considered a good target for immunotherapy and its treatment  has been a key goal in recent years. 
Here we consider a randomized, double-blind, phase III trial whose 
primary objective   was to determine the safety and efficacy of the combination of a melanoma 
immunotherapy (gp100) together with an antibody vaccine (ipilimumab) in patients with previously treated metastatic melanoma (\citealp{Hodi2010}). 
Despite the original endpoint was objective response rate   at week 12, it was amended to overall survival and   then considered secondary endpoint.
A total of 676 patients were randomly assigned to receive ipilimumab plus gp100, ipilimumab alone, or gp100 alone.  The study was designed to have at least $90\%$ power to detect a difference in overall
survival between the ipilimumab-plus-gp100   and gp100-alone groups at a two-sided $\alpha$ level of $0.05$, using a log-rank test. 
Cox proportional-hazards models were used to estimate hazard ratios and to test their significance. 
The results showed that ipilimumab with gp100 improved overall survival as compared with gp100 alone in patients with metastatic melanoma. 
However, the treatment had a delayed  effect and  an overlap between the Kaplan-Meier curves was observed during the   first six months. Hence,  the proportional hazards assumption appeared to be no longer valid, and a different approach would had been advisable. 

\begin{figure}[p!]
	\centering
	\includegraphics[width=0.6\linewidth]{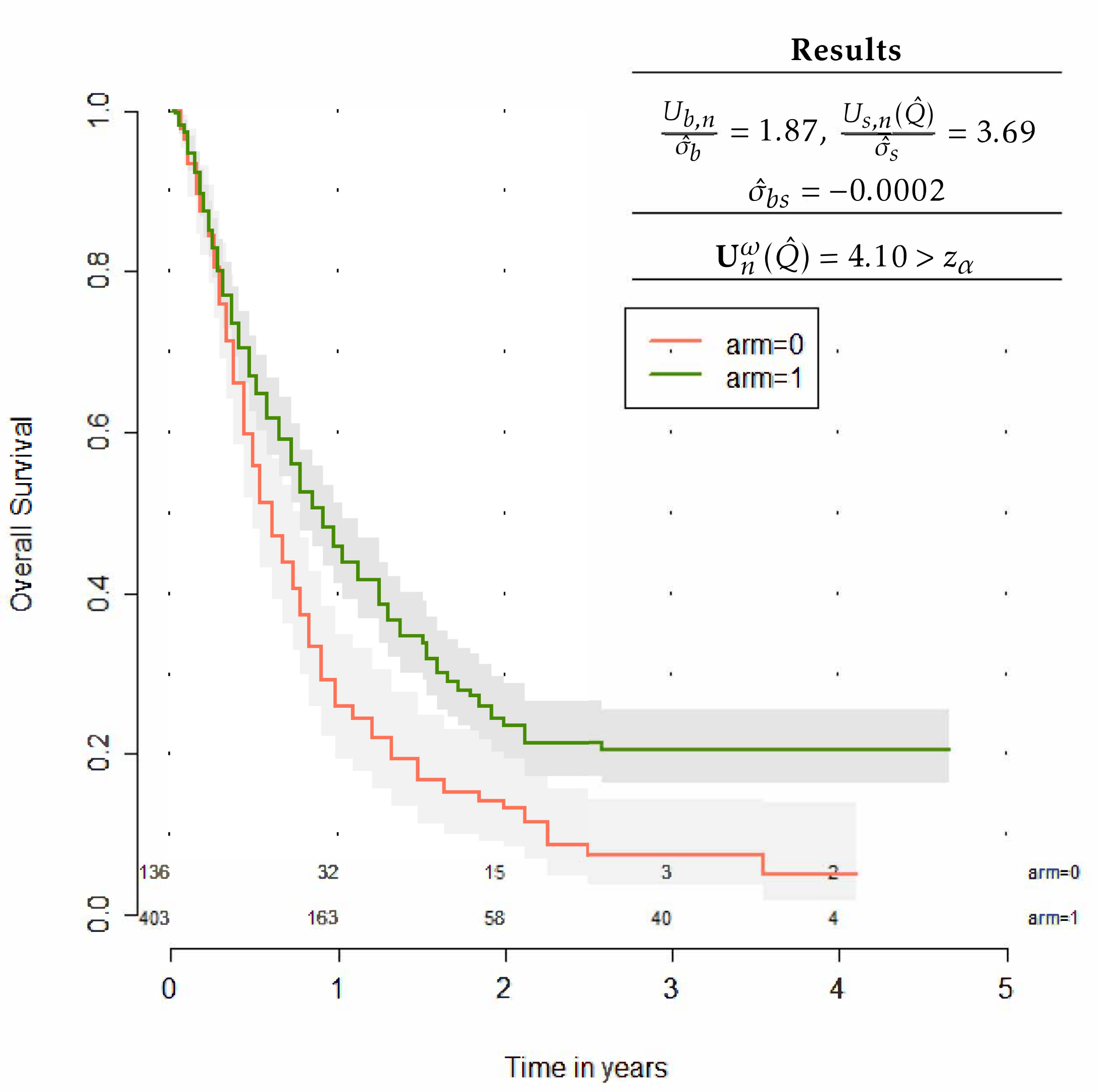}
	\caption{Kaplan-Meier Curves for Overall Survival.}
	\label{fig:figurekm}
\end{figure}

In order to illustrate our proposal, we consider the  comparison  between the ipilimumab-plus-gp100   and   gp100-alone groups based on the overall survival and objective response as co-primary endpoints of the study.
For this purpose,  we have reconstructed individual observed times by scanning the overall survival Kaplan-Meier curves reported in Figure 1A of \citet{Hodi2010} using the \texttt{reconstructKM} package (\citealp{reconstructkm}) (see Figure \ref{fig:figurekm}), and, afterwards, we have simulated the binary response to mimic the percentage of responses obtained in the study.  

Using the data obtained, we employ the $\mathcal{L}$-statistic by means of the function \texttt{lstats} in the \texttt{SurvBin} package. To do so, we need to specify the weights ($\hat{Q},\omega$) to be used, and the time-points ($\tau_0,\tau_b,\tau$).
In our particular case, we take $\tau_0=0,\tau_b=0.5,\tau=4$ according to the trial design,  
choose  $\hat{Q}(t)= \hat{G}(t-)\cdot (1-\hat{S}(t-))$ to account for censoring and delayed effects in late times, and $(\omega_b,\omega_s)= (0.25,0.75)$ to emphasize the importance of overall survival over objective response. The results are summarized in Figure \ref{fig:figurekm}.

Since we obtained $\mathrm{\textbf{U}}_{n}^{\omega}(\hat{Q})= 4.10>z_{\alpha=0.05}$, we have a basis to reject $\mathrm{H}_0$ and conclude that the ipilimumab either improved overall survival or increased the percentages of tumor reduction in patients with  metastatic melanoma, or both.


\section{Simulation study}  

\subsection{Design}

We have conducted a simulation study to  evaluate our proposal in terms of type-I error.
We have generated bivariate binary and time-to-event data  through a
copula-based framework and using conditional sampling as described in  (\citealp{Trivedi2007}).  
The parameters used for the simulation  (summarized in Table \ref{scenarios}) have been the following:
Frank's copula with association parameter $\theta=0.001,2,3$; 
Weibull survival functions, $S_{b,a}(t)=e^{-(t/b)^a}$,  
with $a=0.5,1,2$ and $b=1$; 
probability of having the binary endpoint $p^{(0)}=0.2, 0.4$; and
sample size per arm $n^{(i)}= 1000$.

The censoring distributions between groups were assumed equal and uniform $U(0,c)$ with  $c=1,3$.
Two different follow-up configurations were considered for
$\tau_0 <\tau_b\leq\tau$: (i) $\tau_0=0, \tau_b=0.5, \tau=1$; and
(ii) $\tau_0=0, \tau_b=\tau=1$. 
We have considered the weights: $\hat{Q}(t)=\hat{G}(t-)^\eta \cdot \hat{S}(t-)^\rho \cdot (1-\hat{S}(t-))^\gamma$
with $ \eta,\rho,\gamma =0,1$ and such that $ \eta+\rho+\gamma \geq 1$, 
and $(\omega_b,\omega_s)$ equal to $(0.5,0.5)$. 
For each scenario, we ran 1000 replicates and estimated the significance level ($\alpha=0.05$).

We note that the chosen values of the association parameter $\theta$ correspond to an increasing association between the binary and time-to-event outcomes.  Indeed, the values $\theta=0.001,2,3$ are equivalent to  $0$, $0.32$, $0.45$  in terms of Spearman's rank correlation coefficient between the marginal distributions of the binary and time-to-event outcomes. We have not considered higher values of $\theta$ as they do not fulfill the condition that $S_X^{(i)}(\tau)>0$ ($i=0,1$). 

We have performed all computations using the R software (Version 3.6.2), and on a computer with an Intel(R) Core(TM) i7-6700 CPU, 3.40 GHz, RAM 8.00GB, 64bit operating system. The time required to perform the considered simulations was 52 hours.

\subsection{Size properties} 

\begin{figure}[p!]
	\centering
	\includegraphics[width=0.6\linewidth]{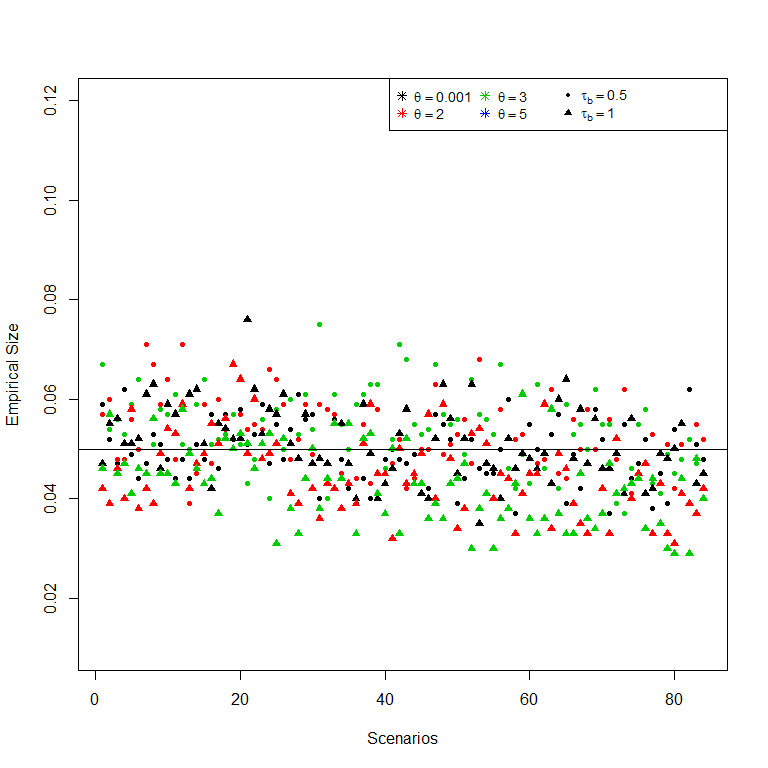}
	\caption{Empirical size according to the $\theta$ and $\tau_b$ parameters.}
	\label{fig:sim}
\end{figure}

The empirical results show that the type I error is very close to the nominal level $\alpha=0.05$ across a broad range of situations. The empirical size resulted in type I errors with a median of 0.049 and  first and third quartiles of 0.043 and 0.055, respectively.
Table \ref{results} summarizes the results according to the parameters of the simulation study. 
The results show that the $\mathcal{L}$-statistics have the
appropriate size and that are not specially influenced by the censoring distribution neither by the selection of weights ($ \eta,\rho,\gamma $).
Figure \ref{fig:sim} displays how the empirical sizes behave according to the association parameter $\theta$ and the follow-up configuration $\tau_b$. We observe that when $\tau_b=1$ the empirical size is slightly small than 0.05. 

We compare the performance of the pooled and unpooled variance estimation and notice that the empirical sizes do not substantially differ between them.


\section{Discussion} 

We have proposed a class of statistics for a two-sample comparison based on two different outcomes: one dichotomous taking care, in most occasions, of short term effects, and a second one addressed to detect long term differences in a survival endpoint. Such statistics test  the equality of proportions and the equality of survival functions. The approach combines a score test for the  difference in proportions and a Weighted Kaplan-Meier test-based for the difference of survival functions.   
The statistics are fully non-parametric and $\alpha$ level for testing the null
hypothesis of no effect on any of these two outcomes.  
The  statistics in the $\mathcal{L}$-class are appealing in situations when both outcomes are relevant, regardless of how the follow-up periods of each outcome are, and even when the hazards are not proportional with respect to the time-to-event outcome or in the presence of delayed treatment effects, albeit the survival curves are supposed not to cross.  
We have incorporated weighted functions in order to control the relative relevance of each outcome and to specify the type of survival differences that may exist between groups.  

The testing procedure using the $\mathcal{L}$-class of statistics satisfies a property called coherence that says that the nonrejection of an intersection hypothesis implies the nonrejection of any sub-hypothesis it implies, i.e.,  $\mathrm{H}_{s,0}$ and $\mathrm{H}_{b,0}$ (\citealp{Romano2005}).  However, the testing procedure based on the $\mathcal{L}$-class of statistics does not fulfill the consonant property that states that the rejection of the global null hypothesis implies the rejection of at least one of its sub-hypothesis. 
\citet{Bittman} faced the problem of how to combine tests into a multiple testing procedure for obtaining a procedure that satisfies the coherence and consonance principles. An extension of this work to obtain a testing procedure that satisfies both properties could be an important research line to consider.

This work has been restricted to those cases in which censoring does not prevent to assess the binary endpoint response. We are currently working on a more general censoring scheme where the binary endpoint could be censored. Last but not least, extensions to sequential  and adaptive procedures in which the binary outcome could be tested at more than one time-point remain open for future research.


\section*{Software} 

The source code to reproduce the results may be found online in the GitHub repository:

\noindent
\url{https://github.com/MartaBofillRoig/SurvBin}.

\section*{Supplementary Material} 

Supplementary Material is available by contacting the authors.

\section*{Acknowledgments}

This work was supported by the Ministerio de Econom\'ia y Competitividad (Spain)  [MTM2015-64465-C2-1-R (MINECO/FEDER)]; the Departament d'Empresa i Coneixement de la Generalitat de Catalunya (Spain) [2017 SGR 622 (GRBIO)]; and the Ministerio de Econom\'{i}a y Competitividad (Spain), through the Mar\'{i}a de Maeztu Programme for Units of Excellence in R\&D [MDM-2014-0445] to M.B.R..  
The authors want to thank Prof. Yu Shen and Prof. Mar\'{i}a Durb\'{a}n for their helpful
comments and suggestions. 

{\it Conflict of Interest}: None declared.

\bibliographystyle{biorefs}

\begin{thebibliography}{} 
	
	
	\bibitem[\protect\citeauthoryear{Alosh, Bretz, and Huque}{2014}]{Alosh2014}
	\textsc{Alosh, M., Bretz, F., and Huque, M.} (2014). Advanced multiplicity adjustment methods in clinical trials. \textit{Statistics in Medicine}, \textbf{33(4)}, 693--713. 
	
	\bibitem[\protect\citeauthoryear{Ananthakrishnan  and Menon}{2013}]{Ananthakrishnan2013} 
	\textsc{Ananthakrishnan, R., and Menon, S.} (2013). Design of oncology clinical trials: A review.
	{\it Critical Reviews in Oncology/Hematology} {\bf 88(1),} 144--153. 
	
	\bibitem[\protect\citeauthoryear{Bauer}{1991}]{Bauer1991}   
	\textsc{Bauer, P.} (1991). Multiple testing in clinical trials. \textit{Statistics in Medicine}. \textbf{10}:871--890. 
	
	\bibitem[\protect\citeauthoryear{Bland and Altman}{1995}]{Bland1995}  
	\textsc{Bland, J. M., and Altman, D. G.} (1995). Multiple significance tests: the Bonferroni method.
	\textit{BMJ}.  \textbf{Jan 21}; 310(6973):170.
	
	\bibitem[\protect\citeauthoryear{Bittman et al.}{2009}]{Bittman}  
	\textsc{Bittman, R. M., Romano, J. P., Vallarino, C., and  Wolf, M.} (2009). Optimal testing of multiple hypotheses with common effect direction. \textit{Biometrika}, \textbf{96(2)}, 399--410. 
	
	\bibitem[\protect\citeauthoryear{Chen and Wang}{2020}]{Chen2020}
	\textsc{Chen, B. E., and Wang, J.} (2020). Joint modeling of binary response and survival for clustered data in clinical trials. \textit{Statistics in Medicine}, \textbf{39(3)}, 326--339. 
	
	\bibitem[\protect\citeauthoryear{De Jong et al.}{2019}]{Jong2019}  
	\textsc{de Jong, W., Aerts, J., Allard, S., Brander, C., Buyze, J., Florence, E., van Gorp, E., Vanham, G., Leal, L., Mothe, B., Thielemans, K., Plana, M., Garcia, F., Gruters, R., and iHIVARNA consortium.} (2019). iHIVARNA phase IIa, a randomized, placebo-controlled, double-blinded trial to evaluate the safety and immunogenicity of iHIVARNA-01 in chronically HIV-infected patients under stable combined antiretroviral therapy. \textit{Trials}. \textbf{20}(1):361. 
	
	\bibitem[\protect\citeauthoryear{Dmitrienko  and Agostino}{2013}]{Dmitrienko2013}
	\textsc{Dmitrienko, A., and Agostino, R. D.} (2013). Traditional multiplicity adjustment methods in clinical trials. \textit{Statistics in Medicine}, \textbf{32(29)}, 5172--5218. 
	
	
	\bibitem[\protect\citeauthoryear{Fleming and Harrington}{1991}]{Fleming1991}  
	\textsc{Fleming, T. R., and Harrington, D. P.} (1991). Counting Processes and Survival Analysis, volume 8. Wiley Online Library.
	
	\bibitem[\protect\citeauthoryear{Gu, Follmann, and Geller}{1999}]{Gu1999}  
	\textsc{Gu, M., Follmann, D., and Geller, N. L.} (1999). Monitoring a general class of two-sample survival statistics with applications. \textit{Biometrika}, \textbf{86(1)}, 45--57. 
	
	\bibitem[\protect\citeauthoryear{Hodi et al.}{2019}]{Hodi2010}   
	\textsc{Hodi, F.S., O'Day, S.J., McDermott, D.F., Weber, R.W., Sosman, J.A., Haanen, J.B., Gonzalez, R., Robert, C., Schadendorf, D., Hassel, J.C., Akerley, W., van den Eertwegh, A.J., Lutzky, J., Lorigan, P., Vaubel, J.M., Linette, G.P., Hogg, D., Ottensmeier, C.H., Lebbé, C., Peschel C., Quirt, I., Clark, J.I., Wolchok, J.D., Weber, J.S., Tian, J., Yellin, M.J., Nichol, G.M., Hoos, A., and Urba, W.J.} (2010). Improved Survival with Ipilimumab in Patients with Metastatic Melanoma. \textit{The New England journal of medicine}. \textbf{363(8)}:711--723. 
	
	\bibitem[\protect\citeauthoryear{Hothorn, Bretz, and Westfall}{2008}]{Hothorn2008}    
	\textsc{Hothorn, T., Bretz, F., and Westfall, P.} (2008). Simultaneous inference in general parametric models. \textit{Biometrical Journal}, \textbf{50(3)}, 346--363.  
	
	\bibitem[\protect\citeauthoryear{Lachin}{1981}]{Lachin1981}  
	\textsc{Lachin, J. M.} (1981). Introduction to Sample Size determination and Power analysis for Clinical Trials. \textit{Controlled Clinical Trials}, \textbf{2}, 92--113.
	
	\bibitem[\protect\citeauthoryear{Lai and Zee}{2015}]{Lai2015}
	\textsc{Lai, X.,  and Zee, B. C. Y.} (2015). Mixed response and time-to-event endpoints for multistage single-arm phase II design. \textit{Trials}, \textbf{16(1)}, 1--10. 
	
	\bibitem[\protect\citeauthoryear{Lai, Lavori, and Shih}{2012}]{Lai2012} 
	\textsc{Lai, T. L., Lavori, P. W., and Shih, M. C.} (2012). Sequential design of phase II-III cancer trials. \textit{Statistics in Medicine}, \textbf{31(18)}, 1944--1960. 
	
	\bibitem[\protect\citeauthoryear{Logan, Klein, and Zhang}{2008}]{Logan2008}  
	\textsc{Logan, B. R., Klein, J. P., and Zhang, M. J.} (2008). Comparing treatments in the presence of crossing survival curves: An application to bone marrow transplantation. \textit{Biometrics}, \textbf{64(3)}, 733--740. 
	
	\bibitem[\protect\citeauthoryear{Lehmann and Romano}{2012}]{LehmannBook}  
	\textsc{Lehmann, E. L., and Romano, J. P.} (2005). Testing Statistical Hypotheses, 3rd ed. New York: Springer
	
	\bibitem[\protect\citeauthoryear{Hess and Gentleman}{2019}]{Hess2019}   
	\textsc{Hess, K., and Gentleman, R.} (2019). R Package 'muhaz': Hazard Function Estimation in Survival Analysis. Version 1.2.6.1.
	
	\bibitem[\protect\citeauthoryear{Mick and Chen}{2015}]{Mick2015}   
	\textsc{Mick, R., and Chen, T.T.} (2015). Statistical Challenges in the Design of Late-Stage Cancer Immunotherapy Studies. \textit{Cancer Immunology Research}, \textbf{3(12)}, 1292--1298. 
	
	\bibitem[\protect\citeauthoryear{Muller and Wang}{1994}]{Muller1994}  
	\textsc{Muller, H.G., and Wang, J.L.} (1994). Hazard Rate Estimation under Random Censoring with Varying Kernels and Bandwidths. \textit{Biometrics}, \textbf{50(1)}, 61--76. 
	
	\bibitem[\protect\citeauthoryear{O'Brien}{1984}]{OBrien1984}
	\textsc{O'Brien, P. C.} (1984). Procedures for comparing samples with multiple endpoints. \textit{Biometrics}, \textbf{04}, 1079--1087.
	
	\bibitem[\protect\citeauthoryear{Papageorgiou  et al.}{2019}]{Papageorgiou2019}		
	\textsc{Papageorgiou, G., Mauff, K., Tomer, A., Rizopoulos, D.} (2019). An Overview of Joint Modeling of Time-to-Event and Longitudinal Outcomes. \textit{Annual Review Of Statistics and Its Application}, \textbf{6(15)}, 1?18.
	
	\bibitem[\protect\citeauthoryear{Pepe and Fleming}{1989}]{Pepe1989}  
	\textsc{Pepe, M. S., and Fleming, T. R.} (1989). Weighted Kaplan-Meier Statistics: A Class of Distance Tests for Censored Survival Data. \textit{Biometrics}, \textbf{45(2)}, 497--507. 
	
	\bibitem[\protect\citeauthoryear{Pepe and Fleming}{1991}]{Pepe1991}   
	\textsc{Pepe, M. S., and Fleming, T. R.} (1991). Weighted Kaplan-Meier Statistics: Large Sample and Optimality Considerations. \textit{Journal of the Royal Statistical Society. Series B (Methodological)}, \textbf{53(2)}, 341--352. 
	
	\bibitem[\protect\citeauthoryear{Pipper, Ritz, and Bisgaard}{2012}]{Pipper2012}  
	\textsc{Pipper, C. B., Ritz, C., and Bisgaard, H.} (2012). A versatile method for confirmatory evaluation of the effects of a covariate in multiple models. \textit{Journal of the Royal Statistical Society. Series C: Applied Statistics}, \textbf{61(2)}, 315--326. 
	
	\bibitem[\protect\citeauthoryear{Pocock, Geller, and Tsiatis}{1987}]{Pocock1987}  
	\textsc{Pocock, S. J., Geller, N. L., and Tsiatis, A. A.} (1987). The analysis of multiple endpoints in clinical trials. \textit{Biometrics}, \textbf{43(3)}, 487--498. 
	
	\bibitem[\protect\citeauthoryear{Romano and Wolf}{2005}]{Romano2005}  
	\textsc{Romano, J. P., and Wolf, M.} (2005). Exact and approximate stepdown methods for multiple hypothesis testing. \textit{Journal of the American Statistical Association}, \textbf{100(469)}, 94--108. 
	
	\bibitem[\protect\citeauthoryear{Rizopoulos}{2012}]{Rizopoulos2012}
	\textsc{Rizopoulos, D.} (2012). Joint Models for Longitudinal and Time-to-Event Data, with Applications in R. Boca Raton: Chapman \& Hall/CRC.
	
	\bibitem[\protect\citeauthoryear{Shen and Fleming}{1997}]{Shen1997}   
	\textsc{Shen, Y., and Fleming, T. R.} (1997). Weighted mean survival test statistics: A class of distance tests for censored survival data. \textit{Journal of the Royal Statistical Society. Series B: Statistical Methodology}, \textbf{59(1)}, 269--280. 
	
	\bibitem[\protect\citeauthoryear{Shen and Cai}{2001}]{Shen2001}  
	\textsc{Shen, Y., and Cai, J.} (2001). Maximum of the weighted Kaplan-Meier tests with application to cancer prevention and screening trials. \textit{Biometrics}. \textbf{57(3)}:837--843. 
	
	\bibitem[\protect\citeauthoryear{Sun}{2020}]{reconstructkm} 
	\textsc{Sun, R.} (2020). R package. GitHub Repository: https://github.com/ryanrsun/reconstructkm 
	
	\bibitem[\protect\citeauthoryear{Thall}{2008}]{Thall2008}  
	\textsc{Thall, P. F.} (2008). A review of phase 2-3 clinical trial designs. {\it Lifetime Data Analysis}, {\bf 14(1)}, 37--53. 
	
	\bibitem[\protect\citeauthoryear{Trivedi and Zimmer}{2007}]{Trivedi2007}
	\textsc{Trivedi, P. K., and Zimmer, D. M.} (2007). Copula modeling: an introduction for practitioners.
	\textit{Foundations and Trends in Econometrics}, \textbf{1(1)}, 1--111.
	
	\bibitem[\protect\citeauthoryear{Tsiatis  and Davidian}{2004}]{Tsiatis2004}	
	\textsc{Tsiatis, A. A., and Davidian, M.} (2004). Joint modeling of longitudinal and time-to-event data: An overview. \textit{Statistica Sinica}, \textbf{14(3)}, 809--834.
	
	\bibitem[\protect\citeauthoryear{Wilson et al.}{2015}]{Wilson2015}
	\textsc{Wilson, M. K., Collyar, D., Chingos, D. T., Friedlander, M., Ho, T. W., Karakasis, K., and Oza, A. M.} (2015). Outcomes and endpoints in cancer trials: Bridging the divide. {\it The Lancet Oncology}, {\bf 16(1)}, e43?e52.
	
\end{thebibliography}


\newpage


\begin{table}[h!]
	\caption{Scenarios simulation study.} \label{scenarios} 
	\begin{tabular*}{\columnwidth}{lc@{\extracolsep{\fill}~\vrule~}lc}
		\hline 	Parameter &  Value &	Parameter &  Value \\ \hline
		$ p^{(0)} $ & $0.2, 0.4$ &
		$ a$ & $ 0.5,1,2 $ \\
		$ b $ & $ 1 $ &
		$ c $ & $ 1, 3 $ \\ 
		$ \theta $ & $0.001, 2, 3$& 	$n^{(i)}$ \hspace{5mm} ($i=0,1$) & $1000$ \\
		$ \tau_b $ & $0.5,1$ &
		$ \tau $ & $1$ \\
		$\omega_b=\omega_s$ & $0.5$  &
		$ \eta,\rho,\gamma $ \hspace{2mm} ($\eta+\rho+\gamma\geq 1 $) & $0,1$  \\ \hline
	\end{tabular*}
	\bigskip
\end{table}

\begin{table}[h!]
	\caption{Median empirical size from $1000$ replications.} \label{results} 
	{\begin{tabular*}{\columnwidth}{@{}l@{\extracolsep{\fill}}c@{\extracolsep{\fill}}c@{\extracolsep{\fill}}c@{}}
			\hline &   & \multicolumn{2}{c}{Variance estimator} \\ \cmidrule{3-4}   
			&  & Pooled & Unpooled  \\ \hline 
			$(\tau_b,\tau)$ & $(0.5,1)$ & 0.052 & 0.050  \\ 
			& $(1,1)$ & 0.046 & 0.047  \\ \hline
			$\theta$ & 0.001 & 0.049 & 0.050  \\ 
			& 2 & 0.048 & 0.048 \\  
			& 3 & 0.049 & 0.048 \\  \hline
			$p^{(0)}$ & 0.2 & 0.049 & 0.049 \\ 
			& 0.4 & 0.049 & 0.048 \\  \hline
			$a$ & 0.5 & 0.052 & 0.050 \\ 
			& 1 & 0.048 & 0.049 \\ 
			& 2 & 0.046 & 0.047 \\  \hline
			$c$ & 1 & 0.049 & 0.049 \\ 
			& 3 & 0.049 & 0.048 \\  \hline 
			$(\rho,\gamma,\eta)$ & (0,1,0) & 0.048 & 0.049 \\ 
			& (1,1,0) & 0.049 & 0.049 \\  
			& (0,0,1) & 0.048 & 0.048 \\ 
			& (0,1,1) & 0.050 & 0.049 \\ 
			& (1,1,1) & 0.048 & 0.050 \\   \hline
	\end{tabular*}}
	\bigskip
\end{table} 


\clearpage 

\end{document}